# Retinal fluorescence microscopy in artificial ophthalmic environment: the Humanized Phantom Eye

**G. Ferraro[1,2], Y. Gigante[1,2], M. Pitea[1,2], L. Mautone[1,3], G. Ruocco[1,4], S. Di angelantonio[1,2,3]\*, M. Leonetti[1,4,5]\***

[1] *Center for Life Nano science @ Sapienza, Istituto Italiano di Tecnologia, Viale Regina Elena, 291, I-00161, Roma, Italy*
[2] *D-Tails, Via di Torre Rossa, 66 - 00165 – Roma*
[3] *Department of Physiology and Pharmacology "V. Erspamer", Sapienza University, 00185 Rome, Italy*
[4] *Dipartimento di Fisica, Università "La Sapienza", Piazzale Aldo Moro, 5, I-00185, Roma, Italy.*
[5] *Soft and Living Matter Laboratory, Institute of Nanotechnology, Consiglio Nazionale delle Ricerche, 00185, Rome, Italy*

*\*corresponding authors: Marco Leonetti email: marco.leonetti@iit.it; Silvia Di Angelantonio email: silvia.diangelantonio@uniroma1.it*

**Abstract:** Phantom eye devices are designed to mimic the optical properties of the human eye; however, the retinal structures are typically emulated with synthetic tissues. Thus, it is currently impossible to exploit phantom eye devices for fluorescence microscopy experiments on cultures, or ex-vivo tissues, although these ophthalmic measurements would be of importance for many applications such as the testing of retinal biomarkers. Here we report and describe the development of a phantom eye designed to host biological samples, such as retinal cultures differentiated from human induced pluripotent stem cells. We characterized the imaging performance of the humanized phantom eye on standard biomarkers such as Alexa Fluor 532 and Alexa fluor 594.

## 1. Introduction

In vivo detection and diagnosis are not always immediately accessible when an instrument or a biomarker is under development or in the early stage of research. Rapid data collection is highly desirable, thus, human phantoms, which are designed to mimic specific tissues or diseases [1–5] are quite common. Generally, phantoms aim to drastically reduce the time and cost of operations offering high sensibility and specificity [6].

In ophthalmology, phantoms are designed to mimic human eye structures and are named phantom eyes (PEs). They can play a pivotal role in ophthalmic device performance validation since they are physical models intended to display similar properties of the human eye, such as optical elements or retinal structures, even if they cannot replace the validation through clinical trials. A considerable amount of literature review has been published on phantom eyes in the last decades. Indeed, a plethora of PE models has been reported in different publications and patents [7]. Some of them are also commercially available, and each of these has specific characteristics depending on the use of the device. For example, PEs are used for ophthalmic training or to calibrate, design, and evaluate the reliability of ophthalmoscopy devices [8–13].

A phantom eye is typically made of a water-tight chamber, which supports: (i) an optical system mimicking the human crystalline lens along with the cornea [10]; (ii) a liquid, usually deionized water (n=1.333), used to replace both the humor vitreous (n = 1.337) and the humor aqueous (n=1.334), where n is the refractive index of the material, and (iii) a spherical fundus.

In addition to these standard elements, this chamber might present other elements used to mimic specific eye parts or characteristics of eye diseases. For example, PEs have been designed and realized to imitate: (i) the anterior chamber[14], (ii) the reflectivity and thickness of the retina including the fovea [3,15], (iii) the optic nerve cup and nerve fiber layers [8,10], and (iv) the retinal detachment and dry age macular degeneration [16]. Other PEs have been designed for specific techniques or to test some instrument reliability such as photoacoustic remote sensing before applying it in vivo on rat retina [17], and to characterize the point spread function of Optical Coherence Tomography (OCT) devices [9].

The eye has always been considered a window to the soul. Nowadays, increasing pieces of evidence demonstrate the presence of retinal manifestations in several diseases typical of the central nervous system. Indeed, as an integral part of the central nervous system, the retina shares many mechanisms with the brain and often mirrors disease progression and hallmarks. Hence, the brain, the spinal cord, and the retina share a common embryological origin and seem to be affected by the same neurodegenerative processes[18,19].

Growing pieces of evidence indicate that ocular disorders exhibit features of neurodegenerative diseases such as Alzheimer's (AD) and Parkinson's disease (PD), showing peculiar alterations and pathological biomarkers at the ocular level[20–22] For this reason, the eye is now considered a window to the brain[23]. Moreover, the eye provides a unique opportunity for direct observation of the nervous tissue thanks to its clear optical media. This pushed interest in using retinal imaging for diagnosis and monitoring neurodegenerative diseases in vivo with noninvasive techniques, bringing with it the need to identify biomarkers to diagnose and monitor neurological conditions.

Thus far, OCT is the standard method to clinically monitor retinal changes [24]. However, in the field of ophthalmology, it is necessary to improve the performance of the existing methodologies for accurate testing and diagnosis.

To optimize a PE resembling the real human eye, we took advantage of the whole mount swine retina and of a newly derived retinal culture obtained through the differentiation of human induced pluripotent stem cells (iPSCs). Indeed, the identification of the molecular factors (Sox2, Klf4, c-Myc, and Oct3/4) able to bring back adult somatic cells to stemness (the "Yamanaka factors") represented a revolution in the field of in vitro human cell-based models [25]. iPSCs are a pluripotent stem cell class able of differentiating, in response to specific stimuli, to any terminally differentiated cell type [26]. Before iPSCs, the development of stem cell-based in vitro models relied mostly on mouse, and in some limited cases on human, embryonic stem cells [27]. Nowadays, the use of human iPSC has pushed the efficient generation of human-based cellular models preventing the ethical issues associated with the use of human embryonic stem cells. Specifically, using a combination of differentiation promoting small molecules, the retinal cell differentiation protocols have been optimized to generate adherent cultures giving retinal cells similar to endogenous human cells [28,29].

In this paper, we present a new phantom eye based on modularity for interchange of biological samples such as human iPSC derived retinal cell cultures, extracted swine retinas, or synthetic samples like fluorescent beads. We studied the influence of fluorescent disks placed behind the biological sample to mimic the natural autofluorescence signal from the retina, and we demonstrated how passing from a flat sample holder to a curved holder enhances the quality of the images limiting the aberrations. We also included in this PE model, named Humanized Phantom Eye (HPE), the possibility to focus the sample along the Z-axis once loaded into a water-tight chamber. Finally, we show images of biological samples on a curved surface to better mimic the human eye properties.

## 2. Methods

*2.1 Humanized phantom eye design and imaging setup*

It is here presented a new generation PE: the Humanized Phantom Eye (HPE). The HPE is capable to host biological tissue of different kinds combined with synthetic elements and fulfills the following characteristics: (i) it is water-tight, (ii) the sample can be focused along the Z-axis once loaded inside, (iii) biological samples or synthetic samples can be easily replaced, (iv) the autofluorescence of the retina can be artificially induced with specifically designed disks, and (v) samples can be placed on flat or curved surfaces.

The HPE, depicted in Fig1a, is composed of an ocular lens (cornea and crystalline) for a total power of 60 diopters, the value used to indicate an emmetropic eye, and a sealed chamber. Within this chamber, Milli-Q water type I (Merck Millipore, Massachusetts, United States), with a refractive index of n=1.333, is loaded to mimic the humor aqueous(n=1.334) thanks to a small chamber between the cornea and the crystalline, but also the humor vitreous (n= 1.337) since it fills the space between the phantom lens and the bottom lid, directly acting as an interface between the sample and the phantom lens. Among the HPE components, we have a custom threaded sample body, that host the sample and allows its movements on the Z-axis, different kind of custom sample holders and disks, and a water-tight bottom with an inlet and an outlet. The inlet and the outlet were designed to easily fill the eye with pure water and to remove the excess air. Particularly the HPE uses as ocular lens, the lens of the commercial PE (OEMI-7, Ocular Instruments, Inc., Bellevue, WA), which has a pupil diameter of 7 mm, total power of 60 diopters, where the cornea and the crystalline are made off poly(methyl methacrylate) (PMMA with n= 1.49) [9]. However, adaptors for other kind of commercial or custom synthetic lens can be easily designed and readily 3d printed.

The external HPE body is realized by custom modular water-tight aluminum threaded parts, except for the bottom lid, that has been thought to be transparent to minimize the signal coming from the surface behind the sample, which could introduce some artifacts in the final image. Therefore, a gorilla glass (Ø 20.00 mm and 1.10 of thickness) with the coating $MgF_2$ (400-700nm) purchased from (Edmund Optics Inc., Barrington, NJ, USA) has been glued on the aluminum threaded component. Instead, the inner parts of the HPE were 3D printed with the Original Prusa i3 MK3S+ (Prusa Research, Prague). Among the 3D printed parts, we have designed two different sample holders, one suitable for round (12 mm) glass coverslips (Thorlabs, Inc., NJ, United States) with a maximum glass thickness of 0.3 mm, and the other suitable for 12.7 mm diameter glass with curved surfaces used to mimic the curvature of the retina. The curved surfaces are made by BK7 Plano-Convex and BK7 Plano-Concave lenses (EKSMA Optics Vilnius, Lithuania) which have been used as an explanted swine retina holder. The Plano-Concave lens has a radius of curvature of 13 mm, the thickness of 4.00+/- 0.20 mm, and a diameter of 12.7 mm. In contrast, the Plano-Convex lens has a radius of curvature of -13 mm, thickness of + 3.69+/-0.20 mm, e diameter of 12.7 mm. Those lenses have a radius of curvature of about 13 mm, the value that approximates the curvature value of the human retina (11-13 mm) [30,31].

The 3D printed holders were designed to allow the air bubbles, trapped into the phantom, to be removed from the imaging area, thanks to specific apertures on the peripherical parts. Plastic sample holders and the disks used to induce the autofluorescence are made of PET-G (Polyethylene terephthalate glycol), whereas in TPU (Thermoplastic Polyurethane) were 3D printed the support for centering the ocular lens (OEMI-7, Ocular Instruments, Inc., Bellevue, WA), and the sealing ring. Both the filaments were purchased from (RS Components Srl., London, United Kingdom). All the 3D-designed models are available on GitHub [32].

As evinced from the exploded model (Fig1a), the sample is loaded within a threaded focusing sample holder body, and it is locked between two threaded rings. The external thread of the body allows to finely adjust the sample position along the Z-direction to better focus the sample. The treaded focusing body screws into the threaded main body which provided the right focusing distance from the phantom ocular lens and the sample, estimated to be ~15.5 mm as reported in the cross-section of the HPE of Fig1c. Fig1b reports a 3D CAD model of the HPE.

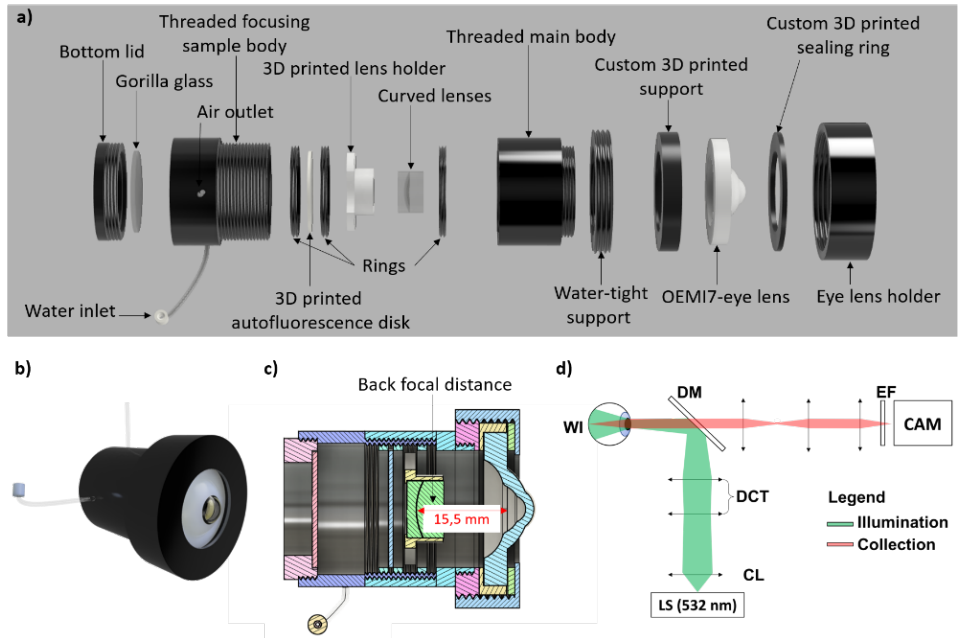

**Fig. 1.** Humanized Phantom Eye (HPE) model. **a)** shows the exploded 3D cad model of the HPE with its components; **b)** shows the compact view of the HPE; **c)** reports the cross-section of it with the estimated back focal distance of the OEMI7-lens; **d)** shows the schematic of the setup used for the imaging, where the acronyms stand for WI (Widefield Illumination), DCT (Divergence Correction Telescope), CL (Collimation Lens), LS (Light Source), DM(Dichroic Mirror), EF(Emission Filter), and CAM (CMOS Camera device).

*2.2 Imaging setup*

Images were acquired in fluorescence mode with a custom setup schematically depicted in Fig1d. The setup can be split into two paths: the illumination path and the collection path. The first has been designed to have the wide field illumination (WI) of the fundus of the eye, using a laser light source (LS) with the emission wavelength of 532nm (OXXIUS, Lannion, France).

The laser beam passes through a collimation lens (CL), and successive, in a divergence correction telescope (DCT) before being reflected from a dichroic mirror (DM) and shining light into the HPE. The second path is the collection path, which is made of a dichroic mirror (DM), a 4f lens system, an emission filter (EF), and the camera (CAM). The camera used for these experiments was a Hamamatsu C11440 (Hamamatsu Photonics K.K., Shizuoka, Japan).

*2.3 Generation of retinal cultures*

Retinal cultures were differentiated from healthy human induced-pluripotent stem cell lines (iPSC) according to a previously published protocol with minor modifications [28,29]. The multi-stage differentiation approach lasts about 30 days, and by using small molecules leads to a homogeneous and well-distributed 2D network of retinal neurons. Human iPSCs (iPSC0028, Sigma-Aldrich, St. Louis, Missouri, Stati Uniti) were dissociated into single cells with 1×Accutase (Merck KGaA, Darmstadt, Germany) and plated on growth factor-reduced Matrigel coated plates (dilution 1:100) at a density of 1000 cells/mm$^2$ in mTeSR Plus with 10 µM Rock Inhibitor (RI; Peprotech, Cranbury, New Jersey, Stati Uniti). The development of retinal neurons can be divided into three stages: the formation of a uniform confluent neuroepithelium-like sheet visible at day 6, the clustering of potentially retinal progenitor macro-islands around day 20 and lastly the generation of ramified retinal neurons at day 30. The retinal progenitor cells were plated onto PLO/Laminin (Sigma-Aldrich) coated round cover glasses (Ø 0.12mm, Thorlabs, Newton, New Jersey, Stati Uniti) at a density of 80.000 cells per glass for subsequent analysis.

The neurogenic basal medium (N2B27w/oA) which consists of 50% DMEM/F12 [1:1], 50% Neurobasal with 1% GlutaMAX Supplement, 0.1% Pen-Strep, 1% NEEA, 1% N2 Supplement, and 2% B27 Supplement w/oA (all from Thermo Fisher Scientific, Waltham, Massachusetts, Stati Uniti) was supplemented with a mix of small molecules at different time points: 1 µM Dorsomorphin and 2.5 µM IDE2 freshly added every day from day 0 to day 6 and 10 mM Nicotinamide (all from Sigma-Aldrich) until day 10 to improve retinal cells differentiation; 25 µM Forskolin (Sigma-Aldrich) was supplemented from day 0 to day 30 in order to promote the retinal progenitor cell proliferation and expansion; 10 µM DAPT (Prepotech) was supplemented from day 20 to day 30 to enhance the genesis of retinal neurons. For future analysis, three differentiations (three biological replicates) were carried out.

*2.4 Dissection procedures to retina extraction*

Retinal tissues were extracted from the dissection of swine eyes, which were collected in the early morning from the local butcher and transported within 4 hours from the enucleation in PBS 1× solution at 4°C (PBS, Thermo Fisher Scientific Inc., Massachusetts, United States). The swine globes were rinsed into PBS solution for a couple of times before being dissected. The anterior segment of the globe was removed through an initial small incision under the cornea, the humor vitreous aspired with a plastic Pasteur pipette (VWR International, Pennsylvania, United States), and to ease the separation of the retina from the eyecup, four cuts were made in the sclera before to flip it. The optic nerve was cut away, and with the help of a little brush, the retina was transferred in a 24-well-dish containing phosphate-buffered saline to clean it. The retinal tissue was sized in small sections before being transferred and sealed into the glass holder. To avoid the contact of the biological tissue with water once loaded into the HPE, the glass holder has been sealed with nail polish (VWR International, Pennsylvania, United States).

*2.5 Immunostaining*

Retinal neurons attached to PLO/Laminin-coated cover glasses were fixed with 4% paraformaldehyde (PFA, Sigma-Aldrich) for 15 minutes at room temperature, washed three times with 1× PBS, permeabilized with PBS containing 0.2% Triton X-100 (Sigma-Aldrich) for 10 minutes and then blocked for 45 minutes at room temperature with a PBS solution containing 0.1% Tween-20 (Sigma-Aldrich) and 5% goat serum (Merck KGaA, Darmstadt, Germany). Cells were incubated overnight at 4°C with the primary antibody TUJ1 (mouse, 1:2000, Covance, MMS-435P, Princeton, New Jersey). The day after, the primary antibody solution was washed out, and the cells were incubated with secondary antibody Goat anti-rabbit Alexa Fluor™ 532 (1:750, Thermo Fisher Scientific Inc., Massachusetts, United States) for 1 hour at room temperature. After primary and secondary antibody staining, washes were performed with PBS solution containing 0.1% Tween-20. The specificity of the staining was tested by performing control experiments in the absence of primary antibody incubation. To prepare cells for the analysis, the coverslips with attached and stained cells were sealed with a second cover glass using PROLONG (Prolong diamond Antifade Montant with DAPI, Invitrogen). DAPI (4′,6-diamidino-2-phenylindole, dihydrochloride, Thermo Fisher Scientific Inc., Massachusetts, United States) was used to stain nuclei. The specificity of the staining was tested by performing control experiments in the absence of primary antibody incubation.

The dissected and extracted retinal sections were fixed with 4% PFA for 15 minutes at room temperature, rinsed in PBS, and incubated with Isolectin GS-IB4 (Alexa fluor 594, 1:200, Invitrogen) at 4°C in PBS 1× overnight. The day after, just before the imaging analysis, the retinal sections were washed three times in PBS 1× and gently placed between two round flat cover glasses (Ø 0.12mm, Thorlabs, Inc., New Jersey, United States) or two BK7 curved lenses (EKSMA Optics Vilnius, Lithuania) before being sealed with the nail polish (VWR International, Pennsylvania, United States) into the relative 3D printed holder. The retinal tissue was sealed between the cover glasses using DAKO (Fluorescent Mounting Medium, Sigma-Aldrich).

**3. Results and discussions**

In order to validate the HPE and demonstrate its potential, we performed several imaging measurements of: (i) human-induced pluripotent stem cells and extracted swine retina loaded into the HPE on flat glass coverslips, as well as the use of fluorescent disks behind the cell cultures to mimic the autofluorescence of the eye; (ii) synthetic samples loaded into flat and curved surfaces, showing also the possibility to use non-biological samples within the HPE; (iii) extracted swine retinas loaded into curved glass holder to better reproduce the fundamental eye properties, increasing the field of view, and, thus, improving the quality of the image.

*3.1 Biological samples and mimicked autofluorescence*

We observed and reported here, representative images of biological samples loaded between two flat coverslips within the HPE. Fig2a shows a real retinal structure observed through the HPE. Specifically, we show on the right panel, a representative image of the network of retinal neurons differentiated from healthy human iPSC (left), labeled with an antibody against TuJ1 (beta tubulin III), a typical neuronal cytoskeletal protein (green). Moreover, on the left panel, we report a representative image of the retinal vasculature captured through the HPE, obtained labeling blood vessels of an extracted swine retina after overnight staining with Isolectin GS-IB4 (red). Note that both images are more focused in the center compared to the edge of the field. That is due to the flatness of the sample that instead needs to be curved for the optical properties of the eye.

In some cases, by shining specific wavelengths of light into the eyes, typically around 488nm, it is possible to observe retinal autofluorescence. This effect is due to several factors[33]. First, as it is well known, the presence of Lipofuscin at the level of the retinal pigment epithelium[34], induces a strong fluoresces with emission wavelength a yellow-orange light [35,36]. Besides, bisretinoid fluorophores, in photoreceptor outer segments, are also one of the primary sources of the natural autofluorescence of the retina that can be elicited with blue light excitation (488 nm) (short-wavelength fundus autofluorescence, SW-AF)[37].

We thus made a further technical implementation of the HPE, to take into account the overall autofluorescence, allowing the insertion, behind the biological sample, of specific disks mimicking the eye autofluorescence and obtaining a diffuse fluorescent signal coming from the sample. Specifically, inserting a disk behind an iPSC derived retinal culture, and selecting different disk thicknesses at 0.30, 1.20 and 2.40 mm, respectively, we were able to increase the background fluorescent signal as depicted in Fig2b-c. Note that the diffuse signal increases with the thickness of the disks used. In this way, our HPE is even more representative of the human eye and can be used to train new ophthalmological devices in different operating conditions.

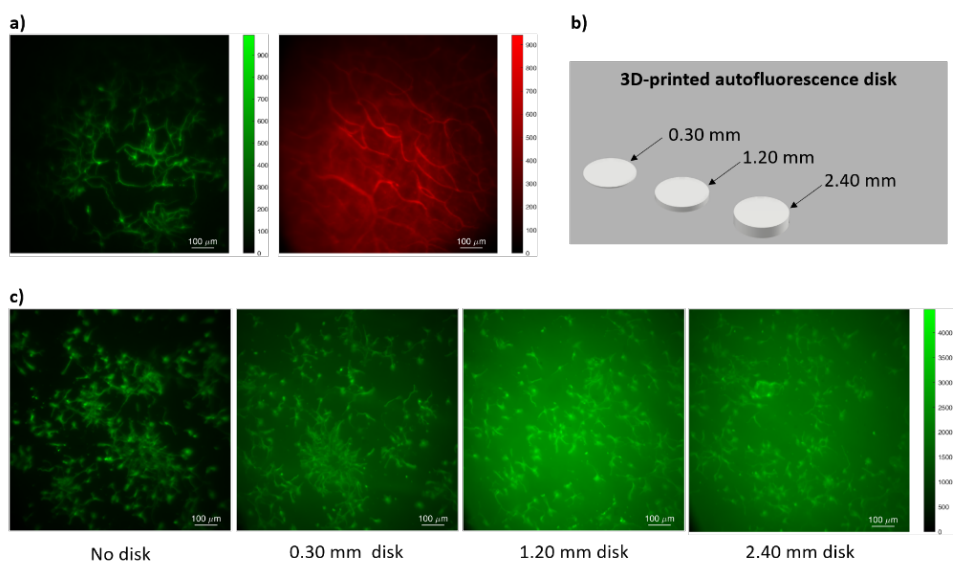

**Fig. 2.** Representative images of biological samples loaded in the HPE. **a)** shows on the left the TUJ1 staining in differentiated iPSC-derived retinal neurons (green) at day 30 of differentiation and on the right the extracted swine retina after the overnight staining with Isolectin GS-IB4 used to label the vascular vessels (red); **b)** shows the 3D printed disks used to mimic the autofluorescence of the retina; **c)** shows different images of retinal neurons (green) using the autofluorescence disks placed behind the biological sample: no disc, 0.30mm, 1.20mm, 2.40 mm of thickness respectively from left to right.

*3.2 Differences in imaging between flat and curved surfaces*

One of the main issues of imaging samples within the HPE is that the samples need to be placed on a curved surface if the aim is to mimic the eye properties. Therefore, as observed before, placing samples between two flat surfaces results in an unfocused and aberrated image at the edge of the field. Thus, by placing samples between curved surfaces, like between two opposite curvature radius lenses (plano-concave/plano-convex lenses), it is possible to improve the result by limiting the aberrations across the field. Here we report on the difference between flat and curved surfaces used within the HPE to host the samples. Specifically, dispersions of ~2µm fluorescent beads (Spherotech Inc, Illinois, United States) diluted in type I water, with a dilution ratio of 100:1, were used to mark the difference between the sample placed within two round glass coverslips and two curved lenses with the curvature radius (R) of R = ± 13mm.

Fig3 shows the differences between images acquired on samples placed on flat and on curved surfaces. Particularly, in Fig3a and Fig3c is depicted a schematic of the sample holder. Fig3b shows on the left the image of the beads and on the right a zoomed area. What is striking in the figure is the difference between beads at the center and at the edge of the field. The beads appear unfocused and with a star shape as we move gradually at the edge of the field, and this effect results more marked by looking into the zoomed area on the right. However, using curved surfaces, as reported in Fig3d, aberration effects at the edge of the field are drastically reduced, as also evinced from the zoomed area on the right.

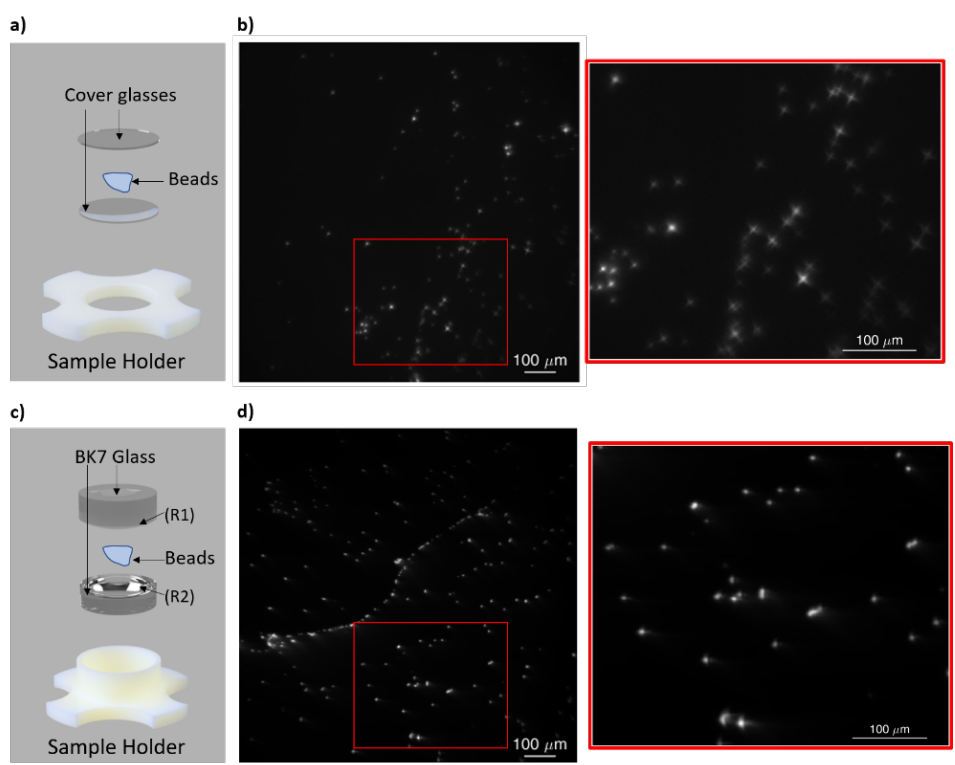

**Fig. 3.** Differences between imaging on a flat and curved glass support. **a)** Shows an illustration of the sample holder and the glass coverslips; **b)** shows the widefield acquisition of the ~2 µm beads loaded between two flat surfaces with the corresponding zoomed area. It is possible to notice how the aberrations rise at the edge of the field due to the flat surfaces; **c)** reports the sample holder illustration for curved surfaces; **d)** shows the widefield acquisition for ~2 µm beads loaded between curved surfaces and the zoomed area respectively. It is noticeable that the aberrations are drastically reduced.

*3.3 Biological samples on curved surfaces*

We then inserted the extracted swine retina between two curved surfaces and acquired within the HPE the fluorescent images of the retinal vasculature, as reported in Fig 4. Samples were prepared from the dissection of the globe of the swine eye following the procedure described in section 2.4, and analyzed after the overnight staining with Isolectin GS-IB4 used to label retinal blood vessels (red). The sample preparation is schematically depicted in Fig4a, whereas acquired images are reported in Fig 4b-c. Fig4b-c show two different pieces of two different retinas, whereas Fig 4c on the right reports a zoomed area of the second retina. What stands out from the zoomed area is that the aberration effects are limited here as also we have seen on beads placed on curved surfaces in Fig3. It is also interesting to see the difference between the retina reported in Fig 2a on the right and those placed instead on curved surfaces (Fig4b-c), where the unfocused areas are drastically reduced.

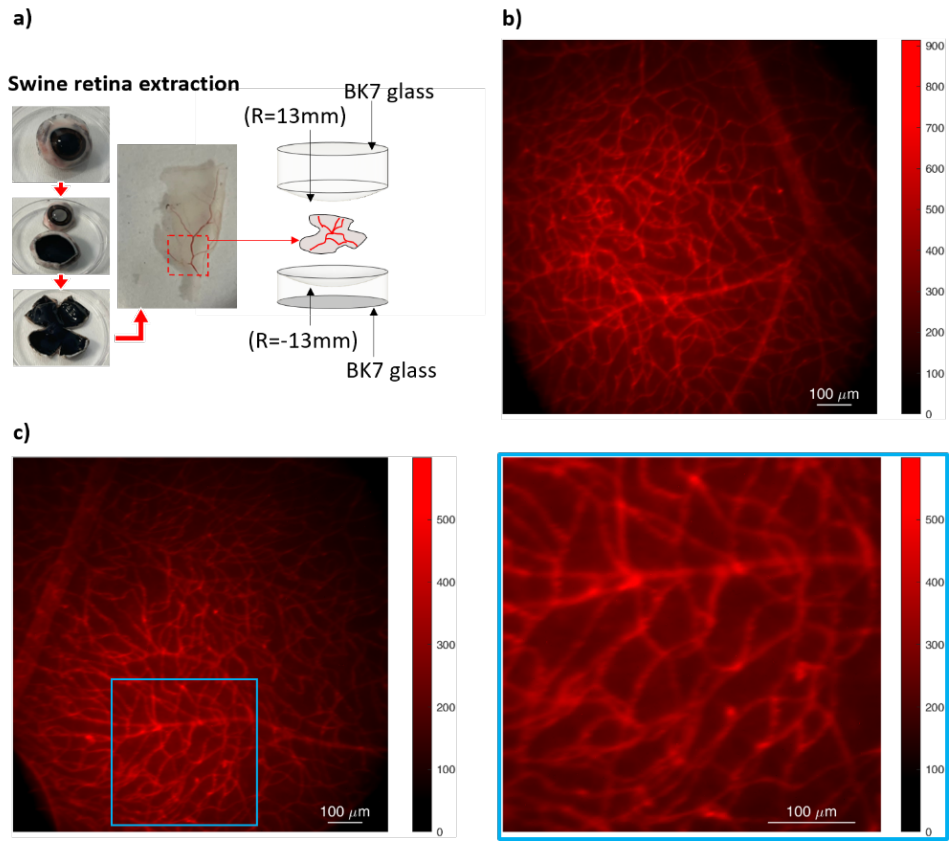

**Fig. 4.** Representative images of extracted swine retina on curved surfaces in the HPE. **a)** shows the main steps of the dissection, retina extraction, and sample loading process; **b-c)** shows wide field images of different extracted swine retinas after the staining with Isolectin GS-IB4 used to label the vascular vessels(red); **c)** on the right reports a zoom of the retina where is possible to notice the strongly reduction of aberrations due to the curvature of the surface hosting the retina.

## 4. Conclusions

A new generation PE has been developed to analyze biological tissues within a physical model designed to mimic the human eye's conditions. We have presented the details and the realization of the HPE as well as imaging characterizations, showing the potential of the device regarding the possibility to use both biological and not-biological target loaded on its fundus such as retinal cells cultures or fluorescent beads.

We have shown the HPE versatility and characteristics: (i) it can be loaded with biological samples such as human retinal cell cultures or extracted swine retinas; (ii) it can be loaded with fluorescent beads for example to enhance the resolution of new devices; (iii) artificial autofluorescence can be induced by placing fluorescent disks on the back of biological samples to increase the diffusion signal to mimic the autofluorescence of the retina and, (iv) samples can be loaded on curved or flat surfaces at the expense of the resolution of the final acquired images.

The use of curved lenses enables to surpass the spherical aberrations visualizing a wider field area of the sample. In contrast, flat coverslips can be used when it might be challenging to load biological samples on a curved surface.

This model of HPE lends itself across many future studies, i.e., it could be loaded with any healthy or unhealthy human or animal tissues to mimic different pathologies reducing the use of in vivo imaging analyses that usually are very expensive, time-consuming, and require ethical validation.

**Acknowledgments**

We would like to thank Dr. Vitantonio Perrone for his valuable contribution to providing our laboratories with enucleated swine globes directly from the local butcher house. The authors thank D-Tails for the realization of the Humanized Phantom Eye (https://www.d-tails.com). M.Leonetti thanks Project LOCALSCENT, Grant PROT. A0375-2020-36549, Call POR-FESR "Gruppi di Ricerca 2020".

**Disclosures**

The authors declare no conflicts of interest.

**Data availability**

Data and Humanized Phantom Eye model underlying the results presented in this paper are not publicly available at this time but may be obtained from the authors upon reasonable request.